\documentclass[submission,copyright,creativecommons]{eptcs}
\providecommand{\event}{WORDS 2011} 
\usepackage{breakurl}             
\usepackage{amssymb,amsmath,amsfonts,amsthm,mathrsfs,srcltx}

\newtheorem{theorem}{Theorem}
\newtheorem{lemma}{Lemma}
\newtheorem{proposition}{Proposition}
\newtheorem{corollary}{Corollary}

\title{Abelian returns in Sturmian words}
\def\titlerunning{Abelian returns in Sturmian words}

\author{Svetlana Puzynina
\institute{University of Turku, Finland}
\institute{Sobolev Institute of Mathematics, Novosibirsk, Russia}
\email{svetlana.puzyina@utu.fi}
\and
Luca Q. Zamboni
\institute{Universit\'e de Lyon, France}
\institute{University of Turku, Finland}
\email{zamboni@math.univ-lyon1.fr}}

\def\authorrunning{S. Puzynina \& L. Q. Zamboni}

\begin{document}
\maketitle

\begin{abstract}
In this paper we study an abelian version of the notion of return
word. Our main result is a new characterization of
Sturmian words via abelian returns. Namely, we prove that a word
is Sturmian if and only if each of its factors has two or three
abelian returns. In addition, we describe the structure of abelian
returns in Sturmian words, and discuss connections between abelian
returns and periodicity.
\end{abstract}

\section{Introduction}

Sturmian words can be defined as infinite words having the
lowest subword complexity among all aperiodic words.
Sturmian words have been widely studied due to their fundamental
importance in different fields of theoretical computer science.
For a survey on some results on Sturmian words we refer to \cite
{lothaire}. Sturmian words have many
equivalent characterizations, e. g. using balanced words, cutting
sequences, mechanical words, and via morphisms. In this paper, we develop the approach based on
the concept of return words.

The notion of a return word is a powerful tool for studying
various problems of combinatorics on words, symbolic dynamical
systems and number theory. Considering each occurrence of a factor
$v$ in an infinite word, the set of return words of $v$ is defined
to be the set of all distinct words beginning with an occurrence
of $v$ and ending just before the next occurrence of $v$. This
notion was introduced by F. Durand and was used for a characterization
of primitive substitutive sequences \cite {durand}. In \cite
{vuillon} it was proved that a word is Sturmian if and only if
each of its factors has two returns; in \cite {jv} the proofs were
simplified and the return words were studied in episturmian words.

In this paper, we establish a similar result for an abelian analogue of
the notion of return word. Two words are abelian equivalent, if
they are permutations of each other. Different abelian properties
of words are widely studied nowadays, such as abelian powers,
avoidance, complexity, abelian periods, etc. We consider return words up to
abelian equivalence: defining abelian returns of a factor $v$ of
an infinite word, we consider all occurrences of factors abelian
equivalent to $v$, and the set of abelian returns is also defined
up to abelian equivalence. As the main result we prove that a word
is Sturmian if and only if each of its factors has two or three
abelian returns. Notice that the methods we used are different
from ones used in \cite {jv, vuillon}.

The paper is organized as follows. After a few preliminary
definitions in Section 2, we discuss in Section 3 connections
between abelian returns and periodicity. In Section 4, we state
our main result concerning characterization of Sturmian words. In
Section 5 we study the structure of abelian returns of Sturmian
words. We prove that every factor of a Sturmian word has two or
three abelian returns; moreover, a factor has two abelian returns
if and only if it is singular. In Section 6 we prove the
sufficiency of the condition on the number of abelian returns for
a word to be Sturmian.

\bigskip

\section{Preliminaries}

We begin by presenting some basics on return words together with key definitions we use in the paper.

Given a finite non-empty set $\Sigma$ (called the alphabet), we
denote by $\Sigma^*$ and $\Sigma^{\omega}$,
respectively, the set of finite words and the set of (right) infinite
words over the alphabet $\Sigma$.
 A word $v$ is a \emph{factor} (resp. a \emph{prefix}, resp. a \emph{suffix}) of a word $w$, if there exist words $x$, $y$ such that $w=xvy$ (resp. $w=vy$, resp. $w=xv$).  The set of factors of a finite or infinite word $w$ is denoted by $F(w)$. Given a finite word $u = u_1 u_2 \dots u_n$ with $n \geq 1$ and
$u_i \in \Sigma$, we denote the length $n$ of $u$ by $|u|$. The
empty word will be denoted by $\varepsilon$ and we set
$|\varepsilon| = 0$. We say that a word $w$ is \emph{periodic}, if
there exists $T$ such that $w_{n+T}=w_n$ for every $n$. A word $w$
is \emph{aperiodic}, if it is not periodic.

Sturmian words can be defined in many different ways. For example,
they are infinite words having the smallest subword complexity
among aperiodic words. The subword complexity of a word is the
function $f(n)$ defined as the number of its factors of length
$n$. For Sturmian words $f(n)=n+1$.

Let $w=w_1w_2\dots$ be an infinite word. The word $w$ is
\emph{recurrent} if each of its factors occurs infinitely many
times in $w.$ In this case, for $u\in F(w)$, let $n_1<n_2<\dots$ be
all integers $n_i$ such that $u=w_{n_i}\dots w_{n_{i}+|u|-1}$.
Then the word $w_{n_i}\dots w_{n_{i+1}-1}$ is a \emph{return word}
(or briefly \emph{return}) of $u$ in $w$. An infinite word
\emph{has $k$ returns}, if each of its factors has $k$ returns.
The following characterization of Sturmian words via return words
was established in \cite {vuillon}:

\begin{theorem}  \label{returns}  {\rm \cite {vuillon}} A recurrent infinite word has
two returns if and only if it is Sturmian. \end{theorem}

Also there exists a simple characterization of periodicity via return words:

\begin{proposition} \label{per} {\rm \cite {vuillon}} A recurrent infinite word is ultimately periodic if and only if there exists a factor having exactly one return word. \end{proposition}

We now define the basic notions for the abelian case. Given a
finite word $u = u_1 u_2 \dots u_n$ with $n \geq 1$ and $u_i \in
\Sigma$, for each $a \in \Sigma$, we let $|u|_a$ denote the number
of occurrences of the letter $a$ in $u$. Two words $u$ and $v$ in
$\Sigma^*$ are \emph{abelian equivalent}
if and only if $|u|_a = |v|_a$ for all $a
\in \Sigma$. We denote it by $u\approx^{ab}v$. It is easy to see that abelian equivalence is indeed
an equivalence relation on $\Sigma^*$.

For an infinite recurrent word $w$ and for $u\in F(w)$, let
$n_1<n_2<\dots$ be all integers $n_i$ such that $w_{n_i}\dots
w_{n_{i}+|u|-1}\approx^{ab} u$. Then the word $w_{n_i}\dots
w_{n_{i+1}-1}$
is an \emph{abelian return word} (or briefly \emph{abelian return}) of $u$ in $w$. We
say that $u$ \emph{has $k$ abelian returns}, if 
the set of its abelian returns consists of $k$ abelian
classes. So, we actually consider abelian classes of returns to
abelian classes.

\medskip

\noindent \textbf{Example.} Consider abelian returns of the factor $01$ of the Thue-Morse word $$t=0110100110010110\dots$$ that is
a fixed point of the morphism $\mu$: $\mu (0)=01$, $\mu (1)=10$.
The abelian class of $01$ consists of two words $01$ and $10$. Consider an occurrence of $01$ starting at position  $i$,
i.e., $t_i=0$, $t_{i+1}=1$. It can be followed by either $0$ or $10$, i.e. we have either $t_{i+2}=0$ or $t_{i+2}=1$, $t_{i+3}=0$.
In the first case we have $t_{i+1}t_{i+2}=10$, which is abelian equivalent to $01$, and hence we have an abelian return $t_{i}=0$. 
In the second case $t_{i+1}t_{i+2}=11$, which is not abelian equivalent to $01$, so we consider the next factor $t_{i+2}t_{i+3}=10\approx^{ab}01$, which gives the abelian return $t_{i}t_{i+1}=01$. 
Symmetrically, $10$ gives abelian returns $1$ and $10$. So, in
total the abelian class of $01$ has three abelian returns: $0$,
$1$ and $01\approx^{ab}10$.
\medskip

In this paper we establish a new characterization of Sturmian
words analogous to Theorem \ref {returns}. Namely, we prove that a
recurrent infinite word is Sturmian
 if and only if each of its factors has two or three abelian returns. On the other hand, contrary to property of being Sturmian, abelian returns do not give a simple characterization of periodicity analogous to Proposition \ref {per}.

\section{Abelian returns and periodicity}

First we prove a simple sufficient condition for periodicity:

\begin{lemma} \label {periodic} Let $|\Sigma|=k.$ If each factor of a recurrent infinite word over the alphabet $\Sigma$
has at most $k$ abelian returns, then the word is
 periodic. \end{lemma}

\noindent\emph{Proof.}  Let $w$ be a recurrent word over a $k$-letter alphabet, and let
$v$ be a factor of $w$ containing
all letters from the alphabet. Consider two occurrences of $v$ in $w$,
say in positions $m$ and $n$ (with $m<n$). Then the abelian class of $w_{m}\dots w_{n-1}$
has all letters as abelian returns, and hence no more, because every factor of $w$
must have at most $k$ abelian returns.
Thus $w$ is periodic with period $n-m$. \qed

\medskip

\noindent \textbf{Remark.} Actually, this proves something
stronger: Let $w$ be any aperiodic word over an alphabet $\Sigma$,
$|\Sigma|=k$, and let $u$ be any factor of $w$ containing $k$
distinct letters, and let $vu$ be any factor of $w$ distinct from
$u$ beginning in $u$. Then the abelian class of $v$ must have at
least $k$ abelian returns. It follows that if a word is not
periodic, then for every positive integer $N$ there exists an
abelian factor of length $>N$ having at least $k+1$ abelian
returns. In other words, the value $k+1$ must be assumed
infinitely often.

\medskip

\noindent \textbf{Remark.} Notice that the condition given by Lemma \ref {periodic} is not necessary for periodicity. It is not difficult to construct a periodic word such that some of its factors have more than $k$ abelian returns.

\bigskip

Notice also that a characterization of periodicity similar to Proposition \ref {per} in terms of abelian returns does not exist. Moreover, in the case of abelian returns it does not hold in both directions. Consider an infinite aperiodic word of the form $\{110010, 110100\}^{\omega}$. It is easy to see that the factor $11$ has one abelian return $110010\approx^{ab}110100$. 
So, the existence of a factor having one abelian return does not
guarantee periodicity. The converse is not true as well: there
exists a periodic word such that each of its factors has at least
two abelian returns. The example is given by the following word
with period 24:
$$w=(001101001011001100110011)^{\omega}.$$
To check that every factor of this word has at least two abelian returns, one can check the factors up to the length $12$. If we denote the period of $w$ by $u$, then every factor $v$ of length $12<l\leq 24$ has the same abelian returns as abelian class of words of length $24-l$ obtained from $u$ by deleting $v$. For a factor of length longer than $24$ its abelian returns coincide with abelian returns of part of this factor obtained by shortening it by $u$.

\section{Characterization of Sturmian words}

The main result of this paper is the following characterization of
Sturmian words:

\begin{theorem} \label {main} An aperiodic recurrent infinite word is Sturmian if and only
if each of its factors has two or three abelian returns.
\end{theorem}

We prove this theorem in the following two sections. The necessity
of the condition on the number of abelian returns is proved in
Section 5, Proposition \ref{necessity}; the sufficiency is proved
in Section 6, Proposition \ref {sufficiency}. Due to space limitations,
we give only a sketch of the proof omitting some of the details. We also
establish some properties of abelian returns of Sturmian words, e.
g., we show that a factor of a Sturmian word has two abelian
returns if and only if it is singular (Section 5, Theorem \ref
{singular}).

\section{The structure of abelian returns of Sturmian words}

In this section we prove the ``only if'' part of Theorem
\ref{main}, and in addition we establish some properties concerning the
structure of abelian returns of Sturmian words.

To describe the abelian returns for Sturmian words, we need to
recall some notation. A factor $u$ of an infinite word $w$ is
called \emph{right special} (\emph{left special}), if $ua$, $ub$
($au$, $bu$) are factors of $w$ for two distinct letters $a$, $b$.
For a Sturmian word there exists exactly one right special factor
of a fixed length. Note also that the set of factors of a Sturmian
word is closed under reversal. A factor is \emph{bispecial}, if it
is right and left special. A factor of a Sturmian word is called
\emph{singular} if it is the only factor in its abelian class.
Notice that singular factors have the form $aBa$, where $a$ is a
letter and $B$ is a bispecial factor. The following proposition
follows directly from definitions and basic properties of Sturmian
words:

\begin{proposition} \label{aBb} Abelian returns of factors of a Sturmian word are either letters or of the form
 $aBb$, where $a\neq b$ are letters, and $B$ is a bispecial factor.
\end{proposition}

\noindent \emph{Proof.} Consider abelian return to a factor $v$ of
length $n$ starting at position $i$. If $w_i=w_{i+n}$, then the
letter $w_i$ is abelian return. If $w_i=a$, $w_{i+n}=b$, $a\neq
b$, then there exists $k\geq 0$, such that $w_{i+1}\dots w_{i+k} =
w_{i+1+n}\dots w_{i+k+n}$, and $w_{i+k+1}\neq w_{i+k+1+n}$. Since
$w$ is balanced, we have that $w_{i+k+1}=b$, $w_{i+k+1+n}=a$. So,
$w_{i+k+2}\dots w_{i+k+n+1}\approx^{ab}v$, and $w_{i}\dots
w_{i+k+1}\approx^{ab} w_{i+n}\dots w_{i+k+n+1}$ is abelian return
to $v$. By definition the factor $w_{i+1}\dots w_{i+k} =
w_{i+1+n}\dots w_{i+k+n}$ is bispecial. \qed

\bigskip

\begin{corollary} \label{1}  In the case of Sturmian words, for each length $l\geq 2$ there exists at
most one abelian return of length $l$. \end{corollary}

\smallskip

Now we proceed to the "only if" part of Theorem \ref{main}:

\begin{proposition} \label {necessity} Each factor of a Sturmian word has two or three abelian returns.
\end{proposition}

The proof of this proposition is based on the characterization of
balanced words presented in \cite {jz}. We will need some notation
from the paper.

Suppose $1\leq p <q$ are positive integers such that $\gcd(p,
q)=1$. Let $\mathscr{W}_{p, q}$ denote the set of all words $w \in
\{0, 1\}^q$ with $|w|_1 =p$. If $w\in \mathscr{W}_{p, q}$ then the
symbol $1$ occurs with frequency $p/q$ in $w$. Define the
\emph{shift} $\sigma: \{0, 1\}^\omega \to \{0, 1\}^\omega$ by
$\sigma(w)_i=w_{i+1}$. Similarly define $\sigma: \{0, 1\}^q \to
\{0, 1\}^q$ by $\sigma(w_0\dots w_{q-1}) = w_1\dots w_{q-1}w_0$.

Since $\gcd(p, q)=1$ then any element of $\mathscr{W}_{p, q}$ has
the least period $q$ under the shift map $\sigma$. We will write
$w\sim w'$ if there exists $0\leq k \leq q-1$ such that $w'
=\sigma^k(w)$. In this case we say that $w$, $w'$ are
\emph{cyclically conjugate}, or that $w$, $w'$ are cyclic shifts
of one another. The equivalence class $\{\sigma^i(w): 0\leq i <
q\}$ of each $w\in\mathscr{W}_{p, q}$ contains exactly $q$
elements. Let
$$\mathbb{W}_{p,q} = \mathscr{W}_{p, q} / \sim$$
denote the corresponding quotient. Elements of $\mathbb{W}_{p,q}$
are called orbits. It will usually be convenient to denote an
equivalence class in $\mathbb{W}_{p,q}$ by one of its elements
$w$.

Given an orbit $[w]\in \mathbb{W}_{p,q}$, let
$$w_{(0)}<_L w_{(1)} <_L \dots <_L w_{(q-1)}$$
denote the lexicographic ordering of its elements. Define the
lexicographic array $A[w]$ of the orbit $[w]$ to be the $q\times
q$ matrix whose $i$th row is $w_{(i)}$. We will index this array
by $0\leq i, j \leq q-1$, so that $A[w] =
(A[w]_{ij})_{i,j=0}^{q-1}$. For $0\leq i, j \leq q-1$,  let
$w_{(i)}[j]$ denote the length-$(j+1)$ prefix of $w_{(i)}$; so the
$w_{(i)}[j]$ are the length-$(j + 1)$ factors of $w$, counted with
multiplicity. For each $j$ this induces the following
lexicographic ordering:
$$w_{(0)}[j] \leq_L w_{(1)}[j] \leq_L \dots \leq_L w_{(q-1)}[ j].$$

\begin{theorem} {\rm \cite{jz}} Suppose $w \in \{0,1\}^q$. The following are equivalent:

\noindent {\rm (1)} $w$ is a balanced word,

\noindent {\rm (2)} $|w(i)[j]|_1\leq |w(i+1)[j]|_1$ for all $0\leq
i\leq q - 2$ and $0\leq j\leq q - 1$.

\end{theorem}

The following proposition from \cite{jz} gives a very practical
way of writing down the lexicographic array associated to a
balanced word.

\begin{proposition} {\rm \cite{jz}} Let $[w]$ be the unique balanced orbit in $\mathbb{W}_{p,q}$. Define $u\in \mathscr{W}_{p, q}$ by
$$u = 0 \dots 0 \underbrace{1 \dots 1}_p$$
Then, for $0\leq i, j \leq q - 1$,

\noindent {\rm (1)} $A[w]_{ij} =(\sigma^{jp}u)_i$,

\noindent {\rm (2)} The $j$th column of $A[w]$ is (the vector
transpose of) the word $\sigma^{jp}u$

\noindent {\rm (3)} $w_{(i)} =u_i(\sigma^{p}u)_i(\sigma^{2p}u)_i
\dots (\sigma^{(q-1)p}u)_i$.

\end{proposition}

\noindent \textbf{Example.} Consider a balanced word $w=0101001\in
\mathscr{W}_{p, q}$. The lexicographic ordering of $[w]$ is
$$0010101 <_L 0100101 <_L 0101001 <_L 0101010 <_L 1001010 <_L 1010010 <_L 1010100,$$
so the corresponding lexicographic array is
$$A[w]= \left( \begin{array}{ccccccc}
0 & 0 & 1 & 0 & 1 & 0 & 1 \\
0 & 1 & 0 & 0 & 1 & 0 & 1 \\
0 & 1 & 0 & 1 & 0 & 0 & 1 \\
0 & 1 & 0 & 1 & 0 & 1 & 0 \\
1 & 0 & 0 & 1 & 0 & 1 & 0 \\
1 & 0 & 1 & 0 & 0 & 1 & 0 \\
1 & 0 & 1 & 0 & 1 & 0 & 0 \end{array} \right) $$

\bigskip

We now apply the above technique for studying abelian
returns as follows:

Fix a Sturmian word $s$ and a factor $v.$ We consider a standard factor $w$ (see, e. g., \cite {lothaire}) of $s$ of long enough length to contain $v$ and all
abelian returns to $v$. Let $|w|=q$, $|w|_1=p$. Then all the
conjugates of $w$ are factors of $s$, they are pairwise distinct,
and $\gcd(p,q)=1$ (see, e. g. \cite{mr}). To be definite, we
assume that $v$ is "poor"\ in $1$-s, i.e., it contains fewer
 $1$'s than the unique other abelian class of the same
length. Then if we consider in $A[w]$ the words $w_{(i)}[j]$, we
have that there exists $n< q-1$ such that
$w_{(i)}[j]\approx^{ab}v$ for $0\leq i \leq n$, and
$w_{(i)}[j]\not\approx^{ab}v$ for $n < i \leq q-1$. Note also that
$A[w]_{im}=A[w]_{(i+q-p)(m+1)}$; from now on the indices are taken
modulo $q$.

The lexicographic array allows to find abelian returns to $v$ in
the following way. For a word $u$ denote by $u[m, l]$ the factor
$u_m \dots u_l$. If for an $i$, $0\leq i \leq n$, we have
$w_{(i)}[k, k+j]\approx^{ab}v$ and $k$ is the minimal such length,
then $w_{(i)}[k-1]$ is abelian return to $v$. Notice also that if
$A[w]_{(i-1)k}=1$ and $A[w]_{ik}=0$, then $w_{(m)}[k,
k+j]\approx^{ab}v$ for $m=i, \dots, i+n$. I. e., we have exactly
$n+1$ words from the abelian class of $v$ starting in every
column, and these words are in consecutive $n+1$ rows (the first
and the last row are considered as consecutive).

\medskip

\noindent\textbf{Example.} Consider abelian returns to the abelian
class of $001$ in the example above. $w_{(i)}[2]\approx^{ab}001$
for $0\leq i \leq 4$; $w_{(i)}[1,3]\approx^{ab}001$ for $i= 4, 5,
6, 0, 1$, $w_{(i)}[2,4]\approx^{ab}001$ for $i= 1, \dots, 5$. So,
the abelian returns are $w_{(0)}[0]=w_{(1)}[0]=0$, $w_{(4)}[0]=1$,
$w_{(2)}[1]=w_{(3)}[1]=01$.

\medskip

\noindent \emph{Proof of Proposition \ref{necessity}.} Suppose
that some factor $v$ of length $j+1$ has $4$ abelian returns, to
be definite let this factor be poor in $1$, and in the
lexicographic array, rows $0\dots n$ start with factors from
the abelian class of $v$. By Corollary \ref {1} there can be at most
one abelian return of a fixed length greater than $1$ (length $1$
will be considered separately), so in a lexicographic array we
have one of the following situations:

\medskip

\noindent 1) there exist $k_1<k_2$ and $n_1<n_2<n$ such that
$w_i[j]$ has abelian returns of length $k_1$ for $i=1,\dots, n_1$,
$w_i[j]$ has abelian returns of length $k_2$ for $i=n_1+1,\dots,
n_2$, and $w_{n_2+1}[j]$ has abelian returns of length greater
than $k_2$;

\smallskip

\noindent 2) symmetric case: there exist $k_1<k_2$ and $n_1<n_2<n$
such that $w_i[j]$ has abelian returns of length $k_2$ for
$i=n_1+1,\dots , n_2$, $w_i[j]$ has abelian returns of length
$k_1$ for $i=n_2+1, \dots , n$, and $w_{n_1}[j]$ has abelian
returns of length greater than $k_2$.

\medskip

We consider case 1) (for case 2) the proof is similar).
First, in case 1) one can notice that the words $w_{n_1}[k_1,
k_1+q]$ and $w_{n_2}[k_2, k_2+q]$ coincide. So if we consider
abelian returns "to the left" of the words $w_{n_1}[k_1, k_1+j]$
and $w_{n_2}[k_2, k_2+j]$, they should be the same, but they are
not: the first one is of length $k_1$, the second one is of length
$k_2$.

It remains to consider the case when $v$ has both letters as
abelian returns. It can be seen directly from the lexicographic
array, that the third and the last return is $01$ (in this case
after a word not from abelian class of $v$ we will necessarily
have a word from abelian class of $v$, i.e., the longest possible
length of abelian return is $2$). \qed

\begin{theorem}\label{singular} A factor of a Sturmian word has two abelian returns if and only if it is singular.
\end{theorem}

\noindent \emph{Proof.} The method of the proof is similar to the
proof of Proposition \ref{necessity} and relies upon the
characterization of balanced words from \cite {jz}.

If a factor is singular, then it is the only word in its abelian
class, so its abelian returns coincide with usual returns. Since
every factor of a Sturmian word has two returns \cite{vuillon},
then a singular factor has two abelian returns.

Now we will prove the converse, i.e., that if a factor $v$,
$|v|=j+1$ of a Sturmian word $s$ has two abelian returns, then it
is singular.

As in the proof of Proposition \ref {necessity}, we consider a
standard factor $w$ of $s$ of long enough length to contain $v$
and all abelian returns to $v$, and denote $|w|=q$, $|w|_1=p$.
Without loss of generality we again assume that $v$ is "poor"\ in
$1$-s, so that there exists $n< q-1$ such that
$w_{(i)}[j]\approx^{ab}v$ for $0\leq i \leq n$, and
$w_{(i)}[j]\not\approx^{ab}v$ for $n < i \leq q-1$.

\medskip

It is not difficult to see that two abelian returns are possible
in one of the following cases:

\medskip

\noindent Case 1) there exist $0 \leq m < n$, $0 < k_1,k_2 < q$
such that $w_{(i)}[k_1-1]$ is abelian return for all $0\leq i\leq
m$, $w_{(i)}[k_2-1]$ is abelian return for all $m+1\leq i \leq n$;

\medskip

\noindent Case 2) there exist $0 \leq m_1 < m_2 < n$, $0< k_1 <
k_2 < q$ such that $w_{(i)}[k_1-1]$ is abelian return for all
$0\leq i\leq m_1$ and $m_2+1\leq i\leq n$; $w_{(i)}[k_2-1]$ is
abelian return for all $m_1+1\leq i \leq m_2$.

\bigskip

\noindent \textbf{Case 1)} In case 1) we will assume that
$k_1< k_2$, the proof in case $k_2< k_1$ is symmetric. We will
consider two subcases:

\medskip

\noindent \textbf{Case 1a)} $A[w]_{m k_2}=1$, $A[w]_{(m+1)
k_2}=0$. This means that $w_{(i)}[k_2, k_2+j]\approx^{ab}v$ for
$i=m+1,\dots, m+n+1$, and $A[w]_{m (k_2-1)}=0$, $A[w]_{(m+1)
(k_2-1)}=1$. 
So, the element $A[w]_{(m+1) k_2}$ is a left-upper element of a
block of abelian class of $v$, and $A[w]_{m (k_2-1)}$ is a
right-lower element of another such block. It is easy to see that
the latter block starts in column $k_1$. Therefore,
$|v|=j+1=k_2-k_1<k_2$.

In case 1a) we will prove that the abelian class of $v$
consists of a single word, i.e., $w_{(i)}[j]=v$ for $i=0,\dots,
n$. Suppose that $w_{(i)}[j]\neq w_{(i+1)}[j]$ for some $i\in
\{0,\dots, n-1\}$. Since the rows grow lexicogaphically, it means
that there exists $0\leq l < j<k_2-1$ such that $A[w]_{il}=0$,
$A[w]_{(i+1)l}=1$. Hence $A[w]_{i(l+1)}=1$, $A[w]_{(i+1)(l+1)}=0$,
and so $w_{(i+1)}[l+1,l+1+j]\approx^{ab} v$. If $m<i+1\leq n$,
then the word $w_{(i+1)}[j]$ has return  $w_{(i+1)}[l]$, which is
impossible, because it has return $w_{(i)}[k_2]$. Similarly we get
that the case  $0\leq i+1\leq m$ and $l+1<k_1$ is impossible.

In  case $0\leq i+1\leq m$ and $k_1\leq l+1 <k_2$ we get that
the word $w_{(i+1)}[k_1,k_1+j]$ has return $w_{(i+1)}[k_1,l]$ of
length $l-k_1+1$. But in this case $w_{(t)}[l+1,
l+1+j]\approx^{ab}v$ for $t=i+1,\dots ,i+1+n$. Contradiction with
the condition that $w_{(t)}[k_2-1]$ is abelian return to
$w_{(t)}[j]$. So, the  case $0\leq i+1\leq m$ and $k_1\leq l+1
<k_2$ is impossible. Hence $w_{(i)}[j] = w_{(i+1)}[j]$ for
$i=0,\dots, n-1$, i.e., the abelian class of $v$ consists of a
single word.

\smallskip

\noindent \textbf{Case 1b)} $A[w]_{m k_2}= 0$ or $A[w]_{(m+1)
k_2}=1$. This means that $w_{(m)}[k_2, k_2+j]\approx^{ab}v$. Hence
the word $w_{(n)}[j]$ has abelian return $w_{(n)}[k_2]$ of length
$k_2+1$, and the word $w_{(m)}[k_1, k_1+j]$ has abelian return
$w_{(m)}[k_1,k_2]$ of length $k_2-k_1+1$, so the returns are
different. This is impossible since $w_{(n)}=w_{(m)}{[k_1,
k_1+q-1]}$.

\bigskip

\noindent \textbf{Case 2)} In case 2) the fact that
$w_{(i)}[k_1]$ is abelian return for all $0\leq i\leq m_1-1$ and
$m_2+1\leq i\leq n$ implies that $n>q/2$. So, $k_1=1$, i.e., we
necessarily have return(s) of length $1$. Since there are two
abelian returns totally, we can have only one return of length
$1$, and this return is $0$. It means that $A[w]_{i0}=0$ for
$0\leq i\leq n$. Since $w_{(m_2)}[1,j+1]\not\approx^{ab} v$ and
$w_{(m_2+1)}[1,j+1]\approx^{ab} v$, we have $A[w]_{m_2 1}=1$,
$A[w]_{(m_2+1)1}=0$, and hence $A[w]_{m_2 0}=0$,
$A[w]_{(m_2+1)0}=1$. We get a contradiction with $A[w]_{i0}=0$ for
$0\leq i\leq n$.

\medskip

So, the converse is proved, i.e., every factor of a Sturmian word
having two abelian returns is singular. \qed

\section{Proof of Theorem \ref {main}: the sufficiency}

Here we prove the "if" part of Theorem \ref {main}, i.e., we
establish the condition on the number of abelian returns forcing
a word to be Sturmian:

\begin{proposition} \label {sufficiency}
If each factor of an aperiodic recurrent infinite word has two or
three abelian returns, then the word is Sturmian.
\end{proposition}

The proof of this proposition is rather technical, it is based on
considering abelian returns to different possible factors of the
infinite word and consecutive restricting the form of the word.
Denote the non-periodic word with $2$ or $3$ abelian returns by
$w$. First we notice that Lemma \ref {periodic} implies that an
aperiodic word with $2$ or $3$ abelian returns must be binary, we
denote letters by $0$ and $1$: $w\in \{0,1\}^{\omega}$. In the
rest of this section instead of abelian returns "to the left" we
consider abelian returns "to the right": if $vu$ is a factor
having $v'\approx^{ab}v$ as its suffix, and $vu$ does not contain
as factors other words abelian equivalent to $v$ besides suffix
and prefix, then $u$ is abelian return to $v$. It is easy to see
that no matter of the definition, the set of abelian returns to
each abelian factor is the same. Though this does not make any
essential difference, this modification of the definition is more
convenient for our proof of this proposition.

\medskip

We say that a letter $a$ is \emph{isolated} in a word
$w\in\Sigma^{\omega}$, if $aa$ is not a factor of $w$. We will
make use of the following key lemma:

\begin{lemma} \label {isolated}
If each factor of an aperiodic recurrent infinite word $w$ has at
most three abelian  returns, then one of the letters is isolated.
\end{lemma}

\noindent\emph{Sketch of proof.} In the proof of this lemma we
will use the following definition. We say that a letter
$a\in\Sigma$ \emph{appears in} $w$ \emph{in a series of length}
$k>0$, if a word $b a^k c$ is factor of $w$ for some letters  $b\neq a$,
$c\neq a$. Considering abelian returns to letters, we get that
every letter can appear in series of at most three different
lengths. Denote these lengths for series of $0$'s by $l_1$, $l_2$,
$l_3$, where $l_1<l_2<l_3$, for series of $1$'s by $j_1$, $j_2$,
$j_3$, where $j_1<j_2<j_3$. Notice that a letter can appear in
series of only two or one lengths, then the third length or the
third and the second lengths are missing.

Consider abelian returns of the word $1 0^{l_1}$: they are $1$, $0^{l-l_1} 1$ for $l=l_2$, $l_3$
(if $0$ appears in series of corresponding lengths), $1^{j-1}
0^{l_1}$ for $j=j_1>1, j_2, j_3$ (if $1$ appears in series of
corresponding lengths) and $0$ for $j_1=1$ . Some of these returns
should be missing or abelian equivalent to others in order to have at most
three abelian returns totally. So we have the following cases:

\smallskip

\noindent -- $j_2$, $j_3$, $l_3$ are missing, i.e., $w \in \{
0^{l_1}1^{j_1}, 0^{l_2}1^{j_1} \}^{\omega}$. In this case abelian
returns are $1$, $0^{l_2-l_1} 1$, and  $1^{j_1-1}0^{l_1}$ for
$j_1>1$ or $0$ for $j_1=1$.

\noindent -- $l_2$, $l_3$, $j_3$ are missing, i.e., $w \in \{
0^{l_1}1^{j_1}, 0^{l_1}1^{j_2} \}^{\omega}$. Abelian returns are
$1$, $1^{j_2-1}0^{l_1}$, and $1^{j_1-1}0^{l_1}$, if $j_1>1$, or
$0$, if $j_1=1$.

\noindent -- $j_2$, $j_3$ are missing, $j_1=2$, $l_2=2l_1$ or
$l_3=2l_1$, i.e., $w \in ( \{ 0^{l_1}, 0^{2l_1}, 0^{l}\} 1^{j_2}
)^{\omega}$. Abelian returns are $1$, $0^{l_1} 1$, $0^{l-l_1} 1$.

\noindent -- $l_3$, $j_3$ are missing, $l_2=2l_1$, $j_1=2$ or
$j_2=2$, $w \in ( \{ 0^{l_1}, 0^{2l_1}\} \{1^{2}, 1^{j}\})^{\omega}$. Abelian returns are $1$,
$0^{l_1} 1$, $1^{j-1}0^{l_1}$ (if $j>1$) or $0$ (if $j=1$).

\smallskip

Notice that the first two cases are symmetric. Considering abelian
returns to the word $1^{j_1}0$, we get symmetric cases ($0$ change
places with $1$, $j_k$ change places with $l_k$, $k=1,2,3$).
Combining the cases obtained by considering abelian returns to $1 0^{l_1}$ with the cases obtained by considering abelian returns to $1^{j_1}0$, we finally get the following remaining cases (up to renaming letters):

\medskip

\noindent 1) $j_2$, $j_3$, $l_3$ are missing, i.e. $w$ is of the
form  $w\in \{ 0^{l_1}1^{j_1}, 0^{l_2}1^{j_1}\}^{\omega}$.

\smallskip

\noindent 2) $l_3$, $j_3$ are missing, $l_1=1$, $l_2=2$, $j_1=2$,
$j_2=4$, i.e. $w\in (\{ 0, 0^2\}\{1^2, 1^4\})^{\omega}$.
\smallskip

\noindent 3) $l_3$, $j_3$ are missing, $l_1=1$, $l_2=2$, $j_1=1$,
$j_2=2$, i.e. $w\in (\{ 0, 0^2\}\{1, 1^2\})^{\omega}$.

\smallskip

\noindent 4) $l_3$, $j_3$ are missing, $l_1=2$, $l_2=4$, $j_1=2$,
$j_2=4$.\, i.e.
 $w\in (\{ 0^2, 0^4\}\{1^2, 1^4\})^{\omega}$.

\medskip

\noindent \textbf{Case 1)}: $w\in \{ 0^{l_1}1^j_1,
0^{l_2}1^j_1\}^{\omega}$.

In the first case we should prove that $j_1=1$. We omit index $1$
for brevity: $j=j_1$. Suppose that $j>1$. Consider abelian returns
to the word $1 0^{l_2}$. They are $1$, $1^{j-1}(0^{l_1}1^j)^k
0^{l_2}$ for all $k\geq 0$ such that the word
$0^{l_2}1^j(0^{l_1}1^j)^k 0^{l_2}$ is a factor of $w$. Therefore,
we have at most two values of $k$ (probably, including $0$).

Abelian returns to the word $1^j 0^{l_1} 1$ are $1$, $(0^{l_2}1^j
)^m 0^{l_1} 1$ for all $m\geq 0$ such that the word
$10^{l_1}1^j(0^{l_2}1^j)^m 0^{l_1} 1$ is a factor of $w$. So, we
have at most two values of $m$ (probably, including $0$).

Taking into account conditions for $m$ and $k$, which we have just
obtained from considering abelian returns to both $1 0^{l_2}$ and
$1^j 0^{l_1} 1$, we find that there are two opportunities for an
aperiodic word $w$:

\smallskip

\noindent \textbf{Case 1a)} $w\in ( \{ (0^{l_1}1^j)^{k_1},
(0^{l_1}1^j)^{k_2} \} 0^{l_2} 1^j)^{\omega}$, $0<k_1<k_2$.
The word $0^{l_2} 1^j 0^{l_1} 1^{j-1}$ has returns $1$,
$0^{l_1}1$, $0^{l_2} ( 1^j 0^{l_1} )^{k-1}1 $ for all $k$ such
that the word $0^{l_2}1^j(0^{l_1}1^j)^k 0^{l_2}$ is a factor of
$w$. To provide at most three abelian returns, $w$ should admit
only one value of $k$. Hence, $w$ is periodic and case 1a) is
impossible.

\smallskip

\noindent \textbf{Case 1b)} $w\in ( 0^{l_1} 1^j, \{ (0^{l_2}1^j)^{m_1},
(0^{l_2}1^j)^{m_2} \} )^{\omega}$, $0<m_1<m_2$.
The word $1^j 0^{l_1} 1^j 0^{l_2} 1$ has returns $1$,
$10^{l_2}$, $10^{l_1} ( 1^j 0^{l_2} )^{m-1} $ for all $m$ such
that the word $10^{l_1}1^j(0^{l_2}1^j)^m 0^{l_1}1$ is a factor of
$w$. To provide at most three abelian returns, $w$ should admit
only one value of $m$. Hence, $w$ is periodic and
case 1b) is impossible.

\smallskip

Thus, in case 1) $1$'s are isolated.

\medskip

\noindent \textbf{Cases 2)--4)} In cases 2)--4) we need to
consider words containing all four series, otherwise we get into
conditions of case 1) in which we proved that $1$-s are isolated.
The proof is similar for the three cases, and is based on studying
abelian returns of certain type. When we examine $w\in (
\{0^{l_1}, 0^{l_2} \}, \{1^{j_1}, 1^{j_2} \})^{\omega}$, we
consider abelian returns to the words $0^{l_1} 1^{j_2}$ and
$0^{l_2} 1^{j_1}$, and with a technical case study obtain that if
both words have at most three abelian returns, then $w$ is
periodic. For brevity, we omit the details of proof for cases
2)--4). \qed

\medskip

\begin{lemma}\label{l1-l2}
If $w\in \{ 0^{l_1}1, 0^{l_2}1 \}^{\omega}$, $0<l_1<l_2$, is an
aperiodic recurrent word and each of its factors has at most three
abelian returns, then $l_2=l_1+1$.
\end{lemma}

\noindent\emph{Proof.} Suppose that $l_2>l_1+1$. Consider abelian
returns to the word $0^{l_1+1}$: it has abelian returns $0$ and
$1(0^{l_1}1)^k 10^{l_1+1}$ for all $k\geq0$ such that $0^l_2 1
(0^{l_1}1)^k  0^{l_2}$ is a factor of $w$, thus there could be at
most two different values of $k$ (probably, including $0$).
Consider abelian returns to the word $10^{l_1}10$: it has abelian
returns $0$ and $(0^{l_2-1}10)^j 0^{l_1-1}1$ for all $j\geq0$ such
that $1 0^{l_1} 1 (0^{l_2}1)^j 0^{l_1} 1$ is a factor of $w$, thus
there could be at most two different values of $k$ (probably,
including $0$). Since $w$ is non-periodic, we have two cases:

\smallskip

\noindent Case I: $w\in (0^{l_2}1 \{ (0^{l_1}1)^{k_1},
(0^{l_1}1)^{k_2} \})^{\omega}$, $0<k_1<k_2$. In this case one can
find four abelian returns to $0^{l_2}10^{l_1-1}$: $0$,
$10^{l_1-1}$, $(10^{l_1})^{k_1-1} 10^{l_2-1}$, $(10^{l_1})^{k_2-1}
10^{l_2-1}$.

\smallskip

\noindent Case II: $w\in (0^{l_1}1 \{ (0^{l_2}1)^{j_1},
(0^{l_2}1)^{j_2} \})^{\omega}$, $0<j_1<j_2$. In this case one can
find four abelian returns to $10^{l_2}10^{l_1}10$: $0$,
$0^{l_2-1}1$, $(0^{l_2-1}10)^{j_1-1} 0^{l_1-1}1$,
$(0^{l_2-1}10)^{j_2-1} 0^{l_1-1}1$. \qed

\bigskip

The proof of Lemma \ref{isolated} and Lemma \ref{l1-l2} imply

\begin{corollary}\label{w} If each factor of an infinite aperiodic recurrent word $w$ has two or three abelian returns, then $w\in \{ 0^{l_1}1, 0^{l_1+1}1 \}^{\omega}$.
\end{corollary}

\begin{lemma}\label{2-balanced}
If each of factors of an aperiodic recurrent infinite word $w$ has
at most three abelian returns, then $w$ is $2$-balanced.
\end{lemma}

\noindent\emph{Proof.} For a length $n$, consider abelian classes
of factors of length $n$ of such word $w$. Denote by $A$ the
abelian class of factors containing the smallest number of $1$-s:
$A=\{ u \in F_n(w): |u|_1=\min_{v\in F_n(w)} |v|_1 \}$. The next
class we denote by $B$: $B=\{ u \in F_n(w): |u|_1=\min_{v\in
F_n(w)} |v|_1+1 \}$, the next one by $C$. If $w$ has only two
abelian classes, then it is Sturmian, so we are interested in the
case when $w$ has at least three abelian classes. For a length
$n$, we associate to a word $w$ a word $\xi^{(n)}$ over the
alphabet of abelian classes of $w$ of length $n$ as follows: for
an abelian class $M$ of words of length $n$, $\xi^{(n)}_k=M$ iff
$w_k\dots w_{k+n-1}\in M$. In other words,
$(\xi^{(n)}_k)_{k\geq0}$ is the sequence of abelian classes of
consecutive factors of length $n$ in $w$.

It is easy to see that $\xi^{(n)}$ contains the following sequence
of classes: $CB^{j_1}A^{j_2}B$ for some $j_1, j_2\geq 1$, i.e.
for some $i$ we have $\xi^{(n)}_i\dots \xi^{(n)}_{i+j_1+j_2+1} =
CB^{j_1}A^{j_2}B$. Then we have
$$
\begin{aligned} &w_i =1, w_{i+n}=0,\\
&w_{k}=w_{k+n} \mbox{ for } k=i+1, \dots, i+j_1-1, \\
&w_{i+j_1} =1, w_{i+j_1+n}=0, \\
&w_{k}=w_{k+n} \mbox{ for } k=i+j_1+1, \dots, i+j_1+j_2, \\
&w_{i+j_1+j_2} =0, w_{i+j_1+j_2+n}=1. \end{aligned}
$$
I. e., $w_i
\dots w_{i+j_1+j_2} = 1 u 1 v 0$, $w_{i+n} \dots w_{i+j_1+j_2+n} =
0 u 0 v 1$.

By Corollary \ref{w} we have $w\in \{ 0^{l_1}1,
0^{l_1+1}1 \}^{\omega}$, so $|u|\geq 2l_1+1$; $u$ contains both
letters $0$ and $1$ and has a suffix $0^{l_1}$. It follows that
$j_2=1$. So, the class $B$ has the following $3$ abelian returns:
$0, 1, 01$. All the returns are of length at most $2$, so if after
an occurrence of $B$ we have $C$, then the next class is $B$
again, otherwise we will get a longer return. So there are no
other classes than these. In addition, we proved that if for
length $n$ there are three abelian classes, then in $\xi^{(n)}$
letters $A$ and $C$ are isolated. \qed

\bigskip

\noindent\emph{Proof of Proposition \ref {sufficiency}.} Due to
Corollary \ref {w} and Lemma \ref {2-balanced}, we have that $w$
is $2$-balanced and it is of the form $\{ 0^{l_1}1, 0^{l_1+1}1
\}^{\omega}$ for some integer $l_1$. Suppose that $w$ is not
$1$-balanced. Then there exists $n$ for which there exist three
classes of abelian equivalence in $F_n(w)$; as above, denote these
classes by $A$, $B$ and $C$. Arguing as in the proof of Lemma \ref
{2-balanced}, consider a sequence of classes $BCB^{j}AB$ which we
necessarily have in $\xi^{(n)}$ for some integer $j$, denote its
starting position by $i-1$. Corresponding factor in $w$ is
$$
\begin{aligned} &w_{i-1}=0, w_{i-1+n}=1, \\
&w_i =1, w_{i+n}=0,\\
&w_{k}=w_{k+n} \mbox{ for } k=i+1, \dots i+j-1, \\
&w_{i+j} =1, w_{i+j+n}=0, \\
&w_{i+j+1}=0, w_{i+j+1+n}=1. \end{aligned}
$$
I. e., $w_i \dots w_{i+j+1} = 1 u 10$, $w_{i+n} \dots w_{i+j+1+n}
= 0 u 01$. Remark that $u=w_{i+1}\dots w_{i+j}$ has prefix
$0^{l_1}10$.

Now consider abelian returns to an abelian class $B0=A1$ of length
$n+1$. The factor starting from the position $i+1$ is of the form
$B0$ so it belongs to this class, and has an abelian return $0$.
The word starting from the position $i+j$ is of the form $B0$ and
has an abelian return $1$. The word starting from the position
$i+l_1-1$ belongs to this class, and has an abelian return $01$.
So we have at least three returns $0$, $1$ and $10$. Now consider
the occurrence of class $B0=A1$ to the left from the position
$i+1$. One can see that the positions $i$ and $i-1$ are from the
class $B1=C0$, so the preceding occurrence of $B0=A1$ has an
abelian return of length greater than $2$, which is a fourth
return, though there should be at most three. So we cannot have
more than two classes of abelian equivalence in an aperiodic word
having two or three abelian returns, i.e., such word should be
$1$-balanced and hence Sturmian. Proposition \ref {sufficiency} is
proved.\qed

\bigskip

\noindent \textbf{Remark.} Actually, in Proposition \ref
{sufficiency} instead of recurrence property one can consider a
weaker property of abelian recurrence in the sense that for every
factor $u$ of $w$ there exists a factor $u'$ from the abelian
class of $u$ which occur infinitely many times in $w$.

\bigskip

\noindent \textbf{Acknowledgements.} The first author is partially
supported by a grant from Magnus Ehrnrooth Foundation and by
Russian Foundation of Basic Research (grants 10-01-00424,
09-01-00244). The second author is partially supported by a grant
from the Academy of Finland and by grant no. 090038011 from the
Icelandic Research Fund.

\bibliographystyle{eptcs}
\bibliography{puzynina}










\end{document}